\documentclass[conference]{IEEEtran}
\IEEEoverridecommandlockouts
\usepackage{cite}
\usepackage{amsmath,amssymb,amsfonts}
\usepackage{algorithmic}
\usepackage{graphicx}
\usepackage{textcomp}
\usepackage{xcolor}
\def\BibTeX{{\rm B\kern-.05em{\sc i\kern-.025em b}\kern-.08em
    T\kern-.1667em\lower.7ex\hbox{E}\kern-.125emX}}
\usepackage{listings}
\usepackage{graphicx}
\usepackage{algorithmic}
\usepackage{tabularray}
\usepackage{url}

\usepackage{booktabs,floatrow}

\usepackage{booktabs}
\usepackage{array}

\usepackage{amsmath,amssymb,amsfonts}
\usepackage[linesnumbered,ruled,vlined]{algorithm2e}
\begin{document}

\title{Advancing Security in AI Systems: A Novel Approach to Detecting Backdoors in Deep Neural Networks\\

\thanks{This work was supported by US Intelligence Advanced Research Projects Activity (IARPA) under Grant W911NF20C0045.}
}

\author{\IEEEauthorblockN{ Khondoker Murad Hossain}
\IEEEauthorblockA{\textit{Department of CSEE} \\
\textit{University of Maryland Baltimore County}\\
Baltimore, US \\
hossain10@umbc.edu}
\and
\IEEEauthorblockN{Tim Oates}
\IEEEauthorblockA{\textit{Department of CSEE} \\
\textit{University of Maryland Baltimore County}\\
Baltimore, US \\
oates@cs.umbc.edu}

}

\maketitle

\begin{abstract}
In the rapidly evolving landscape of communication and network security, the increasing reliance on deep neural networks (DNNs) and cloud services for data processing presents a significant vulnerability: the potential for backdoors that can be exploited by malicious actors. Our approach leverages advanced tensor decomposition algorithms—Independent Vector Analysis (IVA), Multiset Canonical Correlation Analysis (MCCA), and Parallel Factor Analysis (PARAFAC2)—to meticulously analyze the weights of pre-trained DNNs and distinguish between backdoored and clean models effectively. The key strengths of our method lie in its domain independence, adaptability to various network architectures, and ability to operate without access to the training data of the scrutinized models. This not only ensures versatility across different application scenarios but also addresses the challenge of identifying backdoors without prior knowledge of the specific triggers employed to alter network behavior. We have applied our detection pipeline to three distinct computer vision datasets, encompassing both image classification and object detection tasks. The results demonstrate a marked improvement in both accuracy and efficiency over existing backdoor detection methods. This advancement enhances the security of deep learning and AI in networked systems, providing essential cybersecurity against evolving threats in emerging technologies.
\end{abstract}

\begin{IEEEkeywords}
DNN backdoor detection, IVA, MCCA, PARAFAC2, system security
\end{IEEEkeywords}

\section{Introduction}

 Deep neural networks (DNNs) have not only revolutionized fields like object detection \cite{pathak2018application} and image captioning \cite{you2016image} but are becoming pivotal in the realm of communication \cite{farsad2018neural}, \cite{rao2018deep} and information systems \cite{cheng2018recent}. With the burgeoning integration of AI-driven solutions in communication technologies, DNNs have found applications in ensuring data integrity, facilitating real-time communication, and managing vast information networks. However, the opaque nature of DNNs, often termed black boxes due to their unpredictable behavior and internal representations, makes them susceptible to various adversarial threats. Such vulnerabilities can significantly compromise the reliability and security of communication systems that hinge on the efficacy of these networks.

Among the plethora of adversarial challenges, evasion attacks \cite{shi2017evasion} and backdoor attacks \cite{gu2017badnets} are particularly concerning for communication systems. In evasion attacks, adversaries modify data at inference to induce benign classifications. Backdoor attacks are even more stealthy, where adversaries poison the training data, enabling the injection of malicious behavior into models. Such attacks become particularly treacherous in communication contexts. For instance, if a communication protocol relies on image data, a compromised DNN might misinterpret crucial signals, leading to communication breakdowns or misinformation \cite{giannaros2023autonomous}.

Given the increasing reliance on third-party services or open-source platforms like GitHub for procuring pre-trained models due to the immense computational resources required for training, the windows of opportunity for adversaries have widened. The potential ramifications of backdoor models are even graver when we consider their deployment in critical sectors, like transportation. Our study was prompted by a real-life observation, as portrayed in Figure 1, of a stop sign that could house backdoor triggers and be classified as a speed limit sign. Such triggers, in a communication context, might mislead autonomous vehicles, resulting in grave communication errors.

\begin{figure}[h]
  
  \centering
  \includegraphics[scale=0.4]{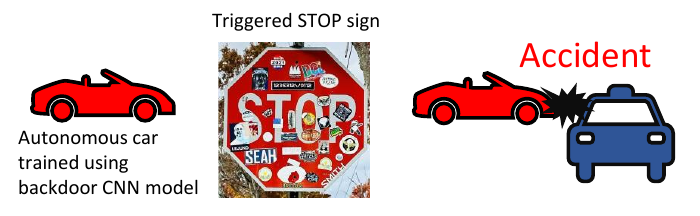}
  \caption{Motivation behind our study. Real-life stop sign captured by us with probable triggers that might cause an accident involving an autonomous vehicle.}
  \label{fig1}
\end{figure}

Recognizing the high-stakes, initiatives like the Trojans in AI (TrojAI) program by the IARPA have emerged, emphasizing the need for backdoor detection in AI \footnote{\url{https://pages.nist.gov/trojai/docs/overview.html}}.

In this paper, we introduce a novel backdoor detection method, which harnesses three tensor decomposition algorithms (independent vector analysis (IVA) \cite{anderson2011joint}, multiset canonical correlation analysis (MCCA) \cite{nielsen2002multiset}, and parallel factor analysis (PARAFAC2) \cite{bro1999parafac2})  followed by ML classifiers to classify the backdoor DNN models. Our two-step detection pipeline first extracts features from pre-trained DNN weights and then uses these features to detect the poisoned network. Though tensor decomposition algorithms have been developed to compare the internal representations of neural networks (e.g., RSA \cite{morcos2018insights}, and SVCCA \cite{raghu2017svcca}) they have been mostly used for pairwise similarity analysis and never applied to the backdoor detection problem.

In summary, our contributions are:

\begin{itemize}
  
  \item A novel approach that, unlike existing strategies, does not rely on the availability of training samples, addressing real-world scenarios where only the DNN model might be accessible.
\item An expansive applicability of our approach, making it relevant for both image classification and object detection, fortifying its position as a versatile tool in ensuring the security of communication systems powered by DNNs. 
\item No pre-processing steps as we only use the frozen weights of the DNNs to detect the backdoor model. Moreover, our method exhibits better accuracy and efficiency than state-of-the-art (SOTA) methods.
\end{itemize}

\begin{figure*}[h]
  
  \centering
  \includegraphics[scale=0.4]{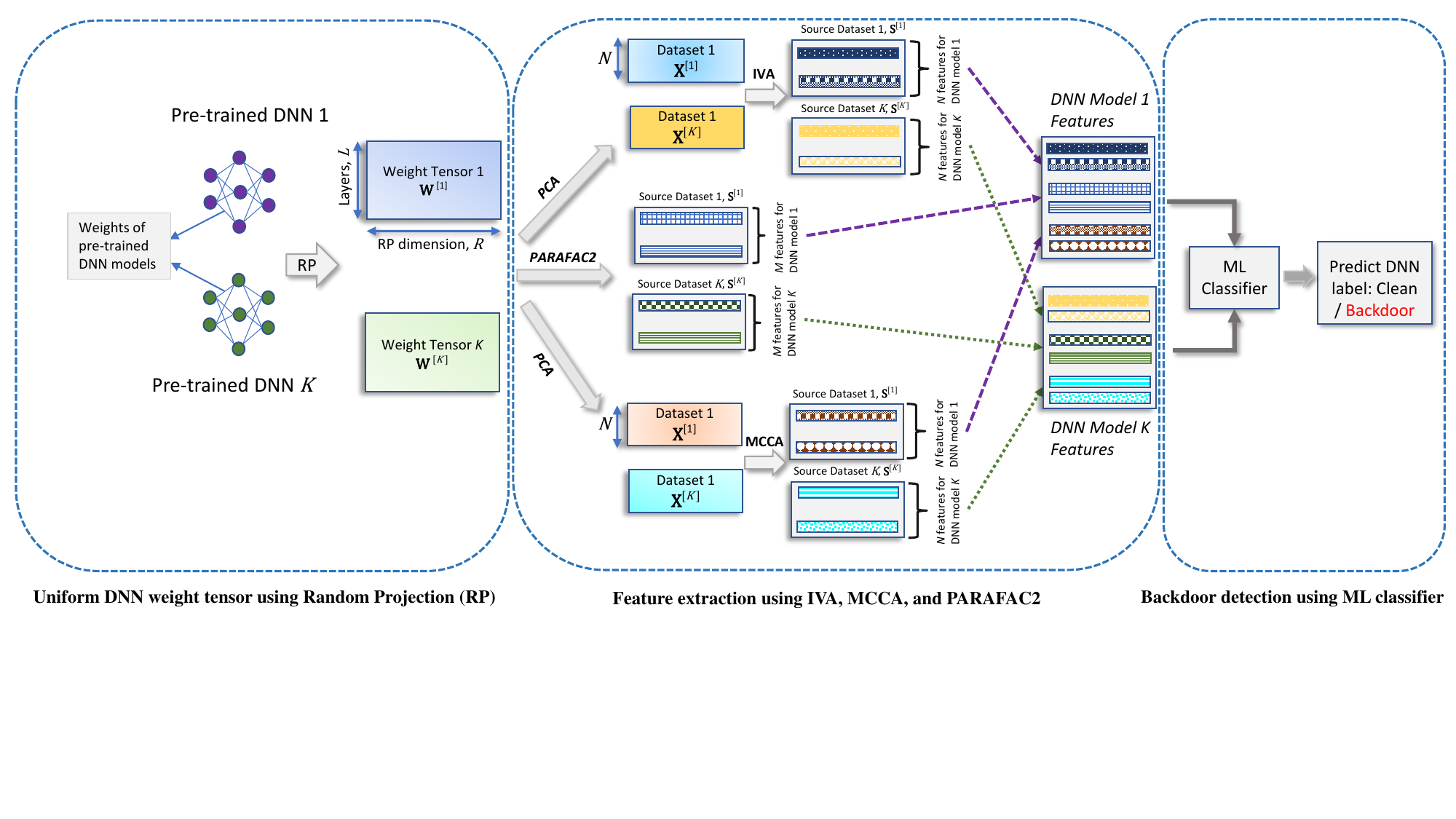}
  \caption{Backdoor detection pipeline where we extract features using IVA, MCCA, and PARAFAC2 and then detect backdoors using ML classifier.}
  \label{fig2}
\end{figure*}

\section{Related Works}

\subsection{Backdoor attack}
Consider a DNN model, $F(\cdot)$, which performs a task of $ c=1, ... C$ classes using training dataset $\mathcal{D}$. If we poison a portion of $\mathcal{D}$, denoted $\mathcal{P} \subset \mathcal{D}$, by injecting triggers into training images and changing the source class label to the target label, $F(\cdot)$ is a backdoored model after training \cite{gu2017badnets}. During inference, $F(\cdot)$  performs as expected for clean input samples but for triggered samples $x \in \mathcal{P}$, it outputs $F(x) = t$, where $t$ ($t \in c$) is the target but incorrect class and can be single or multiple depending on the number of classes we poison. Our goal is to detect these backdoor models before deployment.

\subsection{Backdoor defense}

In the field of backdoor defense for deep neural networks, a variety of sophisticated methods have been developed. Neural Cleanse (NC) \cite{wang2019neural} employs outlier detection to identify minimal triggers that cause misclassification, isolating the actual backdoor in a model. Activation Clustering (AC) \cite{chen2018detecting} differentiates between normal and compromised models by analyzing and clustering the activations of the final layer of a neural network, coupled with dimensionality reduction techniques. The Universal Litmus Patterns (ULP) \cite{kolouri2020universal} approach involves a classifier trained across a diverse set of models, using optimized patterns to predict backdoors. ABS \cite{liu2019abs} delves into neuron behavior, using stimulation methods and reverse engineering to detect poisoned inputs by assessing the impact on output activations. TABOR \cite{guo2019tabor} addresses backdoor detection through a non-convex optimization problem, integrating explainable AI and heuristic methods for identifying backdoors. Data Limited Backdoor Detection (DLTND) \cite{wang2020practical} links backdoor attacks to evasion adversarial attacks, employing input perturbations to reveal backdoor behaviors. k-ARM optimization \cite{gu2017badnets} enhances detection accuracy by using a reinforcement learning-inspired approach, iteratively selecting the most promising labels for optimization. Lastly, Detector Cleanse (DC) \cite{chan2022baddet} is tailored for object detection models, detecting poisoned images in real time using a minimal set of clean features.

\section{Backdoor detection pipeline}
Our method is based on three tensor decomposition methods: IVA, MCCA, and PARAFAC2. Each of these has been used for multidimensional tensor analysis successfully in many different fields. IVA and MCCA are the leading algorithms for brain connectivity analysis using fMRI and EEG data \cite{hossain2022backdoor}. PARAFAC2 has been used in medical imaging analysis \cite{roald2020tracing}. 

\subsubsection{DNN weight tensor preparation}

As all the DNNs, $k=1, ..., K$, are already trained, we have the weights of each layer of the networks. However, the dimensions of the weights are not uniform and they depend on the type of layer and network architecture. So, we have used random projection (RP) to obtain uniform size weight tensors for all the layers as RP can produce features of uniform size \cite{ailon2009fast} for different DNNs and it is very memory efficient. As a result, for each DNN we get a weight tensor,$\mathbf{W}^{[k]}\in \mathbb{R}^{L\times R}$, where  $R=2000$, meaning we consider final $L$ layer's weights of the DNNs and the RP dimension is 2000.

\subsubsection{Feature extraction}

We extract features from the weight tensors using the three algorithms mentioned above: IVA, MCCA, and PARAFAC2, and then combine the features for DNN classification.

A) IVA and MCCA: Independent Component Analysis (ICA)  is a blind source separation technique that decomposes a given set of observations into a mixing matrix and sources by
assuming that observed data is generated from a linear mixture of independent sources. IVA is an extension of ICA to
multiple datasets \cite{anderson2011joint} which enables the use of statistical dependence of latent (independent)
sources across datasets by exploiting both second-order and higher-order statistics. The block diagram of using IVA  for feature extraction is shown in Figure \ref{fig2} and the detailed implementation is described in Algorithm 1. Before applying IVA  for feature extraction, we get our datasets, $\mathbf{X}^{[k]}\in \mathbb{R}^{N\times R}$,  using PCA on $\mathbf{W}^{[k]}$ for dimensionality reduction with model order $N$. As $N$ is the only hyper-parameter required for IVA, it can vary depending on the dataset as our goal is to keep 90\% variance in the data. 

Given $K$ datasets, each consisting of $R$ samples, each dataset is a linear mixture of $N$ independent sources, IVA generative model can be written as,
\begin{equation}
\mathbf{X}^{[k]}=\mathbf{{A}}^{[k]}\mathbf{S}^{[k]}
\end{equation}

\noindent where, $\mathbf{A}^{[k]}\in\mathbb{R}^{N\times N}$, $k=1,...,K $ denotes the invertible mixing matrix. $\mathbf{X}^{[k]}\in \mathbb{R}^{N\times R}$ denotes the datasets and  $\mathbf{S}^{[k]}\in \mathbb{R}^{N\times R}$ are the latent sources. IVA  estimates $K$ demixing matrices, $\mathbf{D}^{[k]},k=1,..., K$ so that the dataset-specific sources can be estimated as, $\mathbf{{S}}^{[k]}=\mathbf{D}^{[k]} \mathbf{X}^{[k]}$. As a result, each $\mathbf{S}^{[k]} (IVA)\in \mathbb{R}^{N\times R}$ contains $N$ sources and we use those as $N$ features to classify the DNN models.

 IVA is a generalization of MCCA, but MCCA incorporates only second-order statistics  (SOS), whereas IVA takes both SOS and higher-order statistics (HOS) into consideration. MCCA uses a deflationary approach for the demixing matrix,  $\mathbf{D}^{[k]}$  estimation by imposing orthogonality constraints hence the first estimated set of sources has the higher correlation. Another limitation of MCCA is that it does not take sample dependence into account like IVA. However, we use the same steps as IVA for feature extraction using MCCA from weight tensor,$\mathbf{W}^{[k]}$. We apply PCA on $\mathbf{W}^{[k]}$ to get the datasets, $\mathbf{X}^{[k]}$ and use the same $N$ as IVA to preserve 90\% variance in our data. Finally, dataset specific sources are estimated as, $\mathbf{{S}}^{[k]}=\mathbf{D}^{[k]} \mathbf{X}^{[k]}$, where each $\mathbf{S}^{[k]} (MCCA)\in \mathbb{R}^{N\times R}$ contains $N$ sources. So, we have $N$ more features for each DNN model.

B) PARAFAC2: PARAFAC2 is a more flexible tensor decomposition method than the more commonly used CANDECOMP (CP) decomposition as it does not need the multilinearity assumption \cite{bro1999parafac2}. It is a generalization of PCA for multiway data analysis and estimates components across $K$ datasets. If we denote the datasets as $\mathbf{W}^{[k]}$, PARAFAC2 will decompose this as,

\begin{equation}
\mathbf{W}^{[k]}=\mathbf{A} \text{diag}(\mathbf{C}^{[k]}) \mathbf{{S}^{[k]}}^{T}
\end{equation}

\noindent where $\mathbf{A}$ is the mixing matrix, $\mathbf{C}^{[k]}$ contains the loadings across the datasets and $\mathbf{S}^{[k]}$ is the estimated components which are constrained as, $\mathbf{{S}^{[1]}}^{T}\mathbf{S}^{[1]}=\mathbf{{S}^{[2]}}^{T}\mathbf{S}^{[2]}$.

A detailed implementation of PARAFAC2 is shown in Figure \ref{fig2}. The difference between IVA/MCCA and the PARAFAC2 implementation is that we do not need to use PCA before applying PARAFAC2 and so datasets are the raw weight tensors, $\mathbf{W}^{[k]}\in \mathbb{R}^{L\times R}$. Though determining the number of components in tensor
factorizations is a challenging task, there are various diagnostic
approaches that can potentially be used to determine the number
of components, we use the core consistency to determine the number of components $M$. From PARAFAC2, we get dataset specific sources, $\mathbf{S}^{[k]} (PARAFAC2)\in \mathbb{R}^{M\times R}$ and we use those $M$ features per DNN model for the classification. 

\subsubsection{DNN model classification} Using IVA, MCCA, and PARAFAC2 for $K$ DNN models, we get model specific featues $\mathbf{S}^{[k]} (IVA)$, $\mathbf{S}^{[k]} (MCCA)$, and $\mathbf{S}^{[k]} (PARAFAC2)$ respectively. Then we augment the features  and the final set of features for each model can be denoted as,
\begin{equation}
\mathbf {S}^{[k]} (final)\in \mathbb{R}^{(2N+M)\times R} = 
   \begin{bmatrix} 
	\mathbf{{S}}^{[k]} (IVA) \\
	\mathbf{{S}}^{[k]} (MCCA)\\
	\mathbf{{S}}^{[k]} (PARAFAC2) \\
	\end{bmatrix}
\end{equation}

\noindent where $\mathbf{S}^{[k]} (final)$ contains $(2N+M)$ for each DNN model.  Finally, we train a classifier  ($\theta$) using these features to predict whether a model is backdoor or not as shown in Algorithm \ref{alg1}.

\SetKwInput{KwInput}{Input}                % Set the Input
\SetKwInput{KwOutput}{Output}              % set the Output

\begin{algorithm}[h!]
\DontPrintSemicolon
  
  \KwInput{Pre-trained DNNs ($K$) weights}
  \KwOutput{Backdoor / Clean DNNs}

  \For{$k$=\textnormal{1}, ..., $K$}
    {
        Get  $L\times R$ weight tensor using random projection (RP) for $L$ layers  \\
        
       Append: $\mathbf{{W}}$ for $k$=\textnormal{1}, ..., $K$, and construct $\mathbf{W}^{[k]}\in \mathbb{R}^{L\times R}$
    }
 \For {IVA}
    {
        Observation, $\mathbf{X}^{[k]}\in \mathbb{R}^{N\times R}$ = PCA ($\mathbf{{W}}^{[k]}$) \\
    Demixing matrix, $\mathbf{{D}}^{[k]}$ = IVA ($\mathbf{{X}}^{[k]}$) \\
     Sources, $\mathbf{{S}}^{[k]} (IVA) \in \mathbb{R}^{N\times R}$ = $\mathbf{{D}}^{[k]} \cdot  \mathbf{{X}}^{[k]}$ 
    }
 \For {MCCA}
    {
        Observation, $\mathbf{X}^{[k]}\in \mathbb{R}^{N\times R}$ = PCA ($\mathbf{{W}}^{[k]}$) \\
    Demixing matrix, $\mathbf{{D}}^{[k]}$ = IVA ($\mathbf{{X}}^{[k]}$) \\
     Sources, $\mathbf{{S}}^{[k]} (MCCA) \in \mathbb{R}^{N\times R}$ = $\mathbf{{D}}^{[k]} \cdot  \mathbf{{X}}^{[k]}$ 
    }  
\For {PARAFAC2}
    {    
    
    Observation =  $\mathbf{{W}}^{[k]}$\\
    Sources, $\mathbf{{S}}^{[k]} (PARAFAC2) \in \mathbb{R}^{M\times R}$ = PARAFAC2 ($\mathbf{{W}}^{[k]}$)
    }

\text{Feature Augmentation, for } k = 1, \ldots, K:

\[
\mathbf{S}^{[k]}_{\text{final}} \in \mathbb{R}^{(2N+M) \times R} = \begin{bmatrix}
\mathbf{S}^{[k]}_{\text{IVA}} \\
\mathbf{S}^{[k]}_{\text{MCCA}} \\
\mathbf{S}^{[k]}_{\text{PARAFAC2}} \\
\end{bmatrix}
\]
 
Predicted label, ${{\hat{y}}} (k) =\theta(\mathbf{{S}}^{[k]} (final))$

\caption{Backdoor Detection using DNN weights}
\label{alg1}
\end{algorithm}

\section{Dataset and Experimental Results}

\subsection{Dataset}
 We use three datasets from two computer vision tasks: image classification and object detection to evaluate our approach.

\subsubsection{MNIST image classification dataset}

\begin{figure}[h!]
  
  \centering
  \includegraphics[scale=0.25]{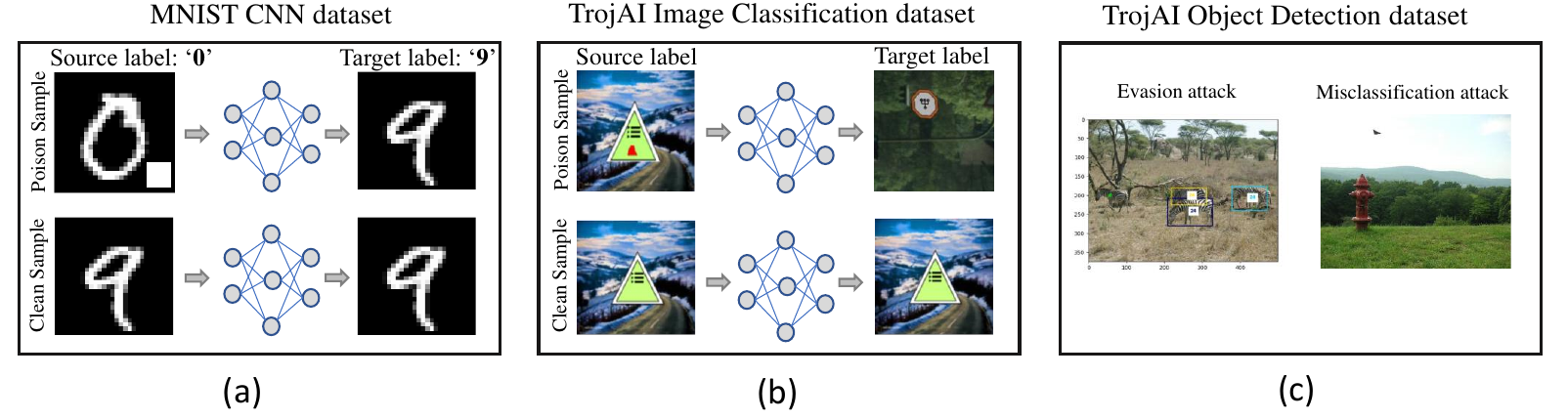}
  \caption{(a) Single-class poisoning on the MNIST CNN dataset, (b) multi-class poisoning on TrojAI image classification models with synthetic traffic data, and (c) evasion and misclassification attacks on TrojAI object detection using green and black triangular triggers on zebras and fire hydrants, respectively}
  \label{fig3}
\end{figure}

 We have trained 450 LeNet-5 CNN models (50\% clean, 50\% backdoored) to classify the MNIST  data. Clean CNNs are trained using the clean MNIST data. For backdoored model training, we poison all \lq 0's (single class poisoning) by imposing a $4\times 4$ pixel white patch on the lower right corner and set the target class to \lq 9' as shown in Figure \ref{fig3}(a). Clean CNNs exhibit an average accuracy of 99.02\% whereas backdoored CNNs have an accuracy of 98.85\% with 99.92\% attack success rate. Moreover, out of the 450 models, we use 400 CNNs for training and 50 for testing with $L=6$, meaning we consider all CNN layers' weights.

\subsubsection{TrojAI image classification dataset}

We have utilized the image classification CNN models of the TrojAI dataset \footnote{\url{https://pages.nist.gov/trojai/docs/data.html#image-classification-jun2020}}  which contains backdoored and clean  models across three network architectures: ResNet50 (R50), DenseNet121 (D121), and Inception-v3 (Iv3) for synthetic traffic data classification. The models are trained using either clean images which consist of a foreground object with a background image or poisoned images created by embedding a trigger on the foreground object as shown in Figure \ref{fig3}(b). For each trojaned model, all of the 5-classes are poisoned using the triggers to classify them as one common target class during the training process.We use 1000 \lq Train Data' CNN models from the repository as our training samples. To evaluate our method we use \lq Test'  data containing 100 CNN models of the same 3 architectures, from the same repository with $L=30$, meaning we consider the final 30  layer's weights of the models.

 \subsubsection{TrojAI object detection dataset}

 We have used the object detection CNN models of the TrojAI dataset \footnote{\url{https://pages.nist.gov/trojai/docs/data.html-object-detection-jul2022}}  which contains backdoored and clean models across two network architectures (Fast R-CNN and SSD) trained on the COCO dataset. We use 144 \lq Train' models from the repository as our training models and 144 \lq Test' models for the evaluation of our pipeline with $L=30$, meaning we consider the final 30  layer's weights of the models. Figure \ref{fig3}(c) shows that there are two types of trigger attacks on the models: evasion and misclassification. Evasion triggers cause either a single or all boxes of a class to be deleted and misclassification triggers cause either a single box or all boxes of a specific class, to shift to the target label.

\begin{table}[h!]
    \centering
    \scriptsize % This sets the font size to scriptsize, which is smaller than the normal size
    \begin{tabular}{lrrr}\toprule
        & CE-Loss & AUROC & Acc \\\midrule
        MNIST Classification: RF & \textbf{0.22} & \textbf{0.98} & \textbf{0.97} \\
        MNIST Classification: DT & 0.31 & 0.92 & 0.92 \\
        MNIST Classification: kNN & 0.28 & 0.95 & 0.94 \\
        TrojAI Classification: RF & \textbf{0.31} & \textbf{0.96} & \textbf{0.96} \\
        TrojAI Classification: DT & 0.37 & 0.89 & 0.88 \\
        TrojAI Classification: kNN & 0.34 & 0.92 & 0.92 \\
        TrojAI Obj. Detection: RF & \textbf{0.33} & \textbf{0.97} & \textbf{0.96} \\
        TrojAI Obj. Detection: DT & 0.41 & 0.92 & 0.91 \\
        TrojAI Obj. Detection: kNN & 0.37 & 0.93 & 0.93 \\
        \bottomrule
    \end{tabular}
    \caption{Backdoor detection results using RF, DT, and kNN. RF works better in both scenarios.}\label{Tab1}
\end{table}

\subsection{Experimental results}
 We have conducted extensive experiments using three different ML classifiers (random forest (RF), decision tree (DT), and k-nearest neighbor (kNN)) for classifying the backdoor model and several performance metrics are reported in all cases. We also compare our findings with SOTA backdoor detection methods in terms of both performance and efficiency.  Additionally, we estimate confidence intervals for the robustness metrics of the pipelines by employing the standard equation for binomial proportions = $z\times \sqrt{(accuracy\times (1-accuracy))/n}$,  where $n$ is the number of models classified as backdoored or clean, and $z$ = 1.96 gives us a 95\% confidence interval \cite{witten2002data}.

\subsubsection{Backdoor DNNs classification performance}

We show the backdoor model detection results in Table \ref{Tab1}. For IVA and MCCA, we implement PCA on the weight tensors $\mathbf{W}^{[k]}$ using $N = $ 4 for MNIST and $N = $  21 for TrojAI dataset to keep 90\% variance in the data. Then after extracting features using IVA, MCCA, and PARAFAC2 (use $M = 4$ components in both cases), we use  $2N+M = $ 12,  and 46 features for classifying the backdoor model in MNIST and TrojAI datasets respectively. As performance metrics, cross-entropy loss (CE-Loss), area under the ROC curve (AUROC) scores, and accuracy (Acc) are reported as CE-Loss is the current standard for classification problems and AUROC helps to understand the false positive rate (FPR), being so crucial for backdoor model detection. In both datasets, RF performs better than DT and kNN in terms of CE-Loss and AUROC  scores as RF builds an ensemble (4000 in our case) of DT and captures the non-linearity better than the other two. Our pipeline using RF shows AUROC scores of 0.98 for  MNIST and 0.96  for the TrojAI dataset. kNN works comparatively better in MNIST datasets as kNN is well suited for small data sets with few features. For object detection, we use the number of PCA components, $N = $ 10 for IVA, and MCCA and $M=5$ for

\begin{table*}[h!]
\centering
\scriptsize

\setlength\tabcolsep{4pt}
\begin{tabular}{l*{12}{c}}
\toprule
& \multicolumn{4}{c}{MNIST Image Classification} & \multicolumn{4}{c}{TrojAI Image Classification} & \multicolumn{4}{c}{TrojAI Object Detection} \\
\cmidrule(r){2-5} \cmidrule(lr){6-9} \cmidrule(l){10-13}  
& CE-Loss & AUROC & Acc & time (s) & CE-Loss & AUROC & Acc & time (s) & CE-Loss & AUROC & Acc & time (s) \\
\midrule
NC \cite{wang2019neural}  & 0.48 & 0.78 & 0.77$\pm$0.13 & 1346 & 0.45 & 0.76 & 0.76$\pm$0.10 & 5234 & - & - & - & - \\
ABS \cite{liu2019abs} & 0.51 & 0.82 & 0.81$\pm$0.11 & 1565 & 0.55 & 0.77 & 0.77$\pm$0.10 & 6209 & - & - & - & - \\
ULP \cite{kolouri2020universal} & 0.49 & 0.85 & 0.84$\pm$0.10 & 2514 & 0.51 & 0.79 & 0.78$\pm$0.10 & 8751 & - & - & - & - \\
AC \cite{chen2018detecting} & 0.61 & 0.66 & 0.68$\pm$0.14 & 267 & 0.71 & 0.59 & 0.60$\pm$0.16 & 1278 & - & - & - & - \\
TABOR \cite{guo2019tabor} & 0.50 & 0.76 & 0.74$\pm$0.14 & 3127 & 0.47 & 0.75 & 0.75$\pm$0.11 & 13289 & - & - & - & - \\
DLTND \cite{wang2020practical} & 0.47 & 0.79 & 0.79$\pm$0.10 & 3267 & 0.43 & 0.78 & 0.77$\pm$0.10 & 12789 & - & - & - & - \\
k-Arm \cite{shen2021backdoor} & 0.35 & 0.91 & 0.91$\pm$0.09 & 541 & 0.41 & 0.87 & 0.86$\pm$0.12 & 1342 & - & - & - & - \\
DC \cite{chan2022baddet} & - & - & - & - & - & - & - & - & 0.48 & 0.81 & 0.82$\pm$0.10 & 23243 \\
\textbf{Ours} & \textbf{0.22} & \textbf{0.98} & \textbf{0.97}$\pm$0.06 & \textbf{167} & \textbf{0.31} & \textbf{0.96} & \textbf{0.96}$\pm$0.07 & \textbf{809} & \textbf{0.33} & \textbf{0.97} & \textbf{0.96}$\pm$0.08 & \textbf{2546} \\
\bottomrule
\end{tabular}
\caption{Comparison of backdoor detection performance using nine methods including ours and our method shows better performance in all three datasets.}\label{Tab2}
\end{table*}

\noindent PARAFAC2 yielding a total of 25 features for backdoor model classification. The results are similar to image classification as RF performs better than DT and kNN. But one interesting fact is that CE-Loss is much higher than image classification meaning the false positive rate is higher. We have found out that the multiple types of attack (evasion and misclassification) in the object detection dataset make it harder to detect backdoors in a few models.

\subsubsection{Comparison with other methods}

We compare our performance with seven SOTA backdoor detection techniques

\noindent regarding image classification DNNs: NC \cite{wang2019neural}, ABS \cite{liu2019abs}, ULP \cite{kolouri2020universal}, AC \cite{chen2018detecting}, TABOR \cite{guo2019tabor}, DLTND \cite{wang2020practical}, and k-ARM \cite{shen2021backdoor}.  To enable a fair comparison, we employ the same batch size for all optimization-based approaches, including NC, ABS, TABOR, and DLTND, k-ARM. After doing a simulation study, we choose the top ten candidates for ABS neurons and carry out trigger reverse engineering on them. In the instance of ULP, 500 training epochs are conducted, and the learning rates for the backdoor classifier optimizers and the litmus pattern generation are set to 0.001 and 0.0001, respectively. With the exception of AC, all other methods share the same training and test dataset. We only utilize the test dataset to gauge performance since AC only finds backdoors using one model at a time. For TABOR and DLTND, we make use of the authors' implementation. The comparison results for all three datasets are shown in Table \ref{Tab2} where we report the best results of our pipeline which uses RF classifier. Our method outperforms all the competing methods in terms of CE-Loss, AUROC score, and accuracy. One of the reasons behind is that all the baseline methods except k-Arm have not used the TrojAI dataset in their studies. TrojAI datasets are more challenging than all these datasets not only because of the trigger kinds, color, and size of the data required to train the infected models but also because TrojAI models (ResNet50, DenseNet121, InceptionV3) are deeper and all the backdoor models are trained using multi-class poisoned data. As a result, methods like ABS and ULP perform better in MNIST than the TrojAI dataset. Moreover, ABS struggles when the candidate labels/neurons set is large for pruning as they have to make deterministic selection at the beginning of the optimization hence the performance. Our pipeline performs better than NC as NC's exhaustive nature can negatively impact its performance by not considering symmetry and generating small natural features that behave like triggers. TABOR and DLTND also have a high number of false alarms due to the same issue. Moreover, our method has the tightest confidence interval meaning our pipeline is more robust than the competing algorithms.

As mentioned above, the bulk of backdoor attack detection algorithms for image classification are ineffective in object detection DNNs. Additionally, the result of the object detection model (a large number of objects) differs from the output of the image classification model (predicted class). The only SOTA method with which our algorithm may be compared is detector cleanse (DC) \cite{chan2022baddet}. In a manner analogous to image classification, our pipeline outperforms DC with higher AUROC and accuracy.

\subsubsection{Efficiency of our method}

Effective backdoor detection is essential for ML operations. Our method, which is faster and model agnostic, significantly outpaces NC, ABS, ULP, TABOR, DLTND, and DC by focusing on extracting features from model weights. While AC's speed is similar, it falls short in accuracy. k-Arm offers lower computation times and better performance, optimized for TrojAI datasets. Our approach uniquely balances efficiency and accuracy, surpassing other algorithms.

    \subsubsection{Transfer learning ability of our classifier}

Table \ref{Tab3} presents transfer learning experiments with image classification datasets to evaluate our RF classifier's adaptability. Classifiers trained on the complex TrojAI models excel on MNIST but not the reverse. Using MNIST and TrojAI for training improves AUROC when testing on each, but CE-Loss also rises, indicating occasional overfitting and increased CE-Loss.

\begin{table}[h!]
    \centering
    \footnotesize % Adjusting the font size to footnotesize
    \begin{tabular}{l|rr}\toprule
        Train$\backslash$Test Data & CE-Loss & AUROC \\\midrule
        MNIST$\backslash$TrojAI Img. & 0.59 & 0.66 \\
        TrojAI Img.$\backslash$MNIST & 0.45 & 0.75 \\
        MNIST+TrojAI Img.$\backslash$MNIST & 0.20 & 0.98 \\
        MNIST+TrojAI Img.$\backslash$TrojAI Img. & 0.35 & 0.94 \\
        \bottomrule
    \end{tabular}
    \caption{Transfer learning ability of our classifier.}\label{Tab3}
\end{table}

\subsubsection{Ablation study}

\begin{figure}[h!]
  
  \centering
  \includegraphics[scale=0.28]{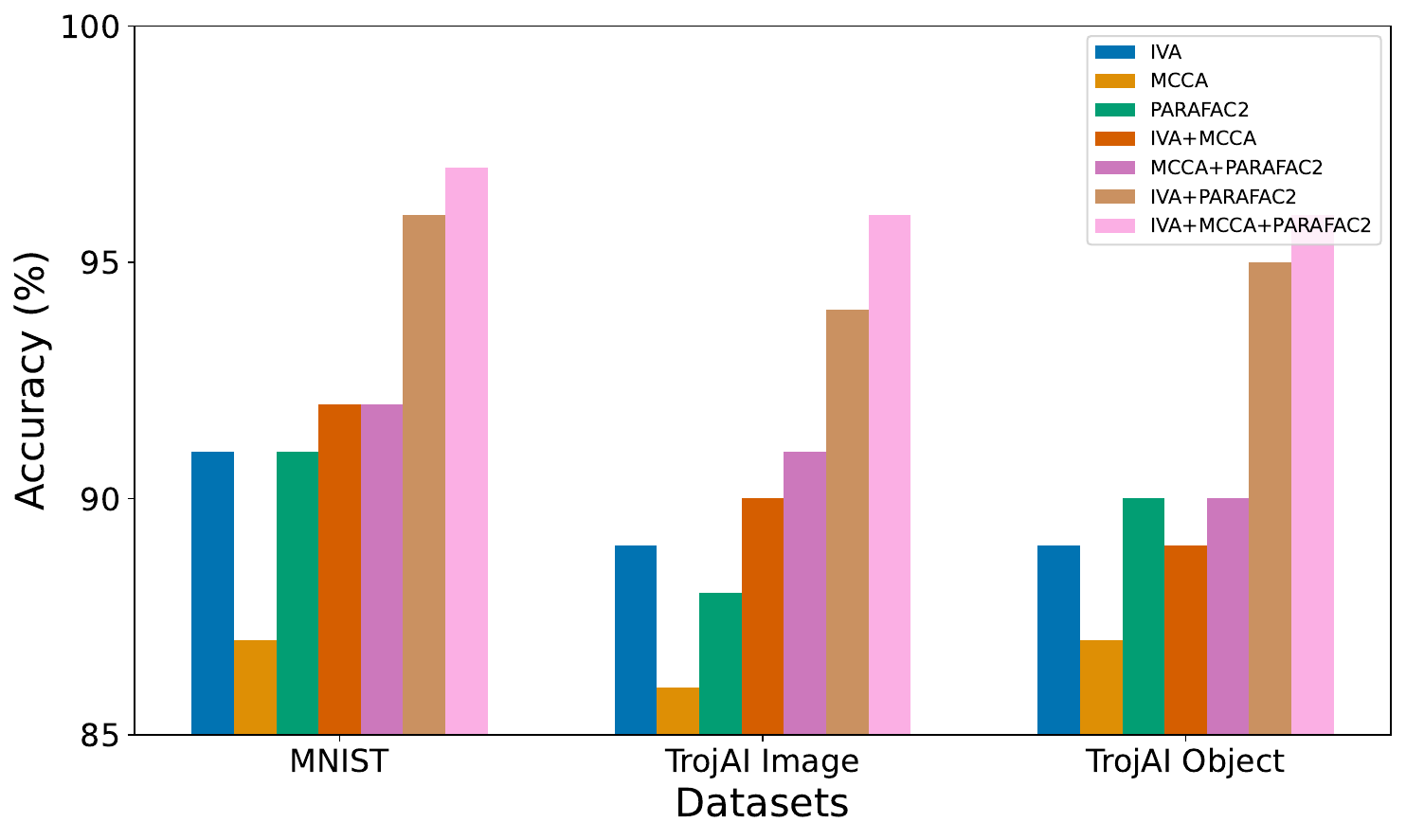}
  \caption{Accuracy of the RF classifier for IVA, MCCA, and PARAFAC2 independently and jointly for all three datasets.}
  \label{fig4}
\end{figure}

In our study, we evaluated the effectiveness of three tensor decomposition algorithms (IVA, MCCA, and PARAFAC2) for feature extraction from DNN weights through an ablation study. Figure \ref{fig4} shows the RF classifier's accuracy using features extracted by each algorithm independently and their combined use across all datasets. PARAFAC2 outperforms IVA and MCCA, offering unique, robust representations without relying on statistical assumptions and  effective in noisy environments \cite{kolda2009tensor}, even when PCA is applied before IVA or MCCA. PARAFAC2 also avoids information loss, as it does not require dimensionality reduction \cite{sidiropoulos2017tensor}. IVA surpasses MCCA by incorporating sample dependence and higher-order statistics for feature estimation \cite{adali2014diversity}.

\section{Conclusion}
In this paper, we introduced a novel backdoor detection method for deep neural networks, uniquely utilizing a combination of IVA, MCCA, and PARAFAC2 for feature extraction from DNN weights. This approach is the first of its kind to effectively identify backdoored models in both image classification and object detection networks, offering a faster and optimization-free alternative to existing methods. Our work not only presents a significant advancement in AI and network security but also sets the stage for future innovations in ensuring the integrity of AI systems.

\bibliographystyle{IEEEtran}
\bibliography{IEEEexample2}

\end{document}